\documentstyle[11pt,IAUS212,twoside,epsf]{article}

\begin{document}

\title{The very rich massive post-main-sequence star population of the
open cluster Westerlund\,1}
 \author{Ignacio Negueruela}
\affil{Departamento de F\'{\i}sica, Ingenier\'{\i}a de Sistemas y
Teor\'{\i}a de la Se\~{n}al, Universidad de Alicante, Apdo. de Correos 99,
E-03080 Alicante, Spain}
\author{J. Simon Clark}
\affil{Department of Physics \& Astronomy, University College London,
Gower Street, London, WC1E\,6BT, England, UK}

\begin{abstract}
We report the discovery of a population of Wolf-Rayet stars in the
young Galactic open cluster Westerlund\,1. In an incomplete shallow
spectroscopic survey, we find six nitrogen-rich (WN) and five carbon-rich
(WC) WR stars. We also confirm the presence of a large population of
yellow supergiants, some of which are candidate hypergiants. Given
this population, Westerlund\,1 is likely to be one of the more massive
young clusters in the Local Group.
\end{abstract}

\vspace*{-2mm}
\section{Introduction}
\vspace*{-2mm}

The highly reddened young open cluster Westerlund\,1 (henceforth Wd\,1) was
reported by Westerlund (1987) to contain a number of both early and
late-type supergiants and some other massive transitional objects.  Exact
determination of its parameters was not possible, but Westerlund (1987)
estimated $d$\,$\simeq$\,5\,kpc and an extinction $A_V$\,$\simeq$\,11\,mag.
Recently, radio continuum observations of Wd\,1 revealed that a number of
cluster members appeared to be associated with very bright radio sources
(Clark \ea 1998; Dougherty, Waters \& Clark, in preparation).  In view of this
result, we carried out a spectroscopic survey of the brighter cluster
members using the Boller \& Chivens spectrograph on the ESO 1.52-m
telescope at La Silla Observatory, Chile.  Low resolution spectroscopy over
the $\sim$\,$\lambda\lambda$\,6\,000\,-\,11\,000\AA\ range was obtained on
the nights of June 24\,-\,26, 2001.

Since the field is very crowded, in addition to the relatively bright
objects that were originally targeted, a large number of fainter stars
was observed. Inspection of the spectra revealed the presence of a
population of faint objects with strong emission lines.

\vspace*{-2mm}
\section{The Wolf-Rayet population}
\vspace*{-2mm}

Despite the low S/N of many of the spectra, due to the faintness of
the objects (the apparently brightest WR stars have $V$\,$>$\,17), it is
immediately possible to identify both nitrogen rich WN (six objects) and
carbon rich WC (five objects) stars. The spectra of all these objects are
displayed in Figures\,1 and 2.


\begin{figure}[t]
\vspace*{20mm}
\plotfiddle{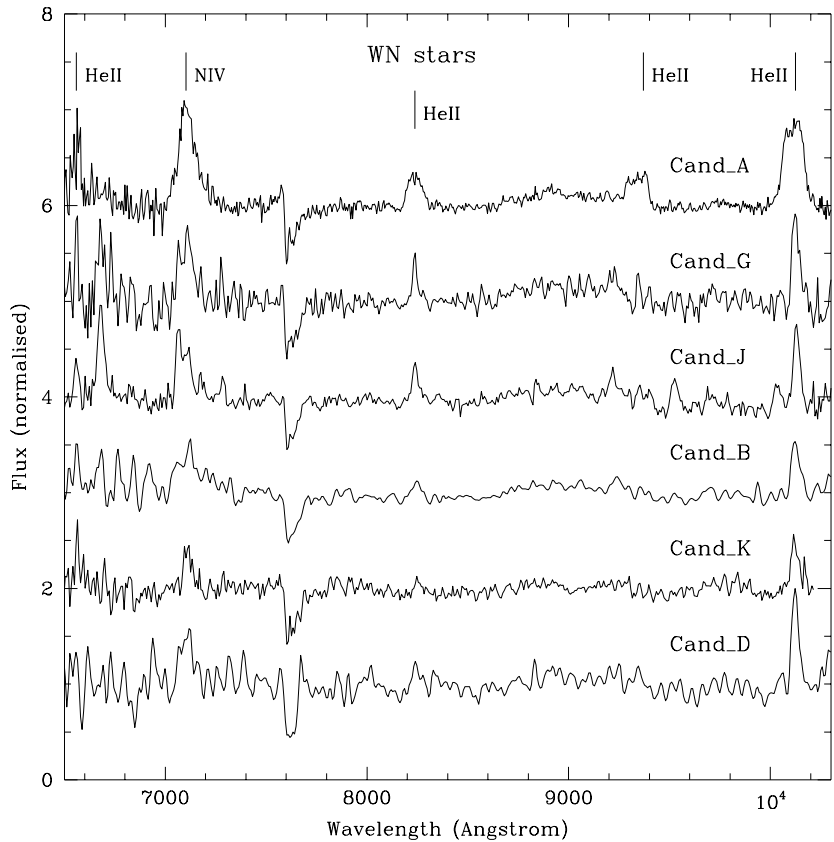}{9cm}{0}{60}{60}{-200}{-100}
\caption{Spectra of the newly discovered WN candidates in Wd\,1,
with prominent transitions identified.}
\end{figure}


\begin{figure}[t]
\plotfiddle{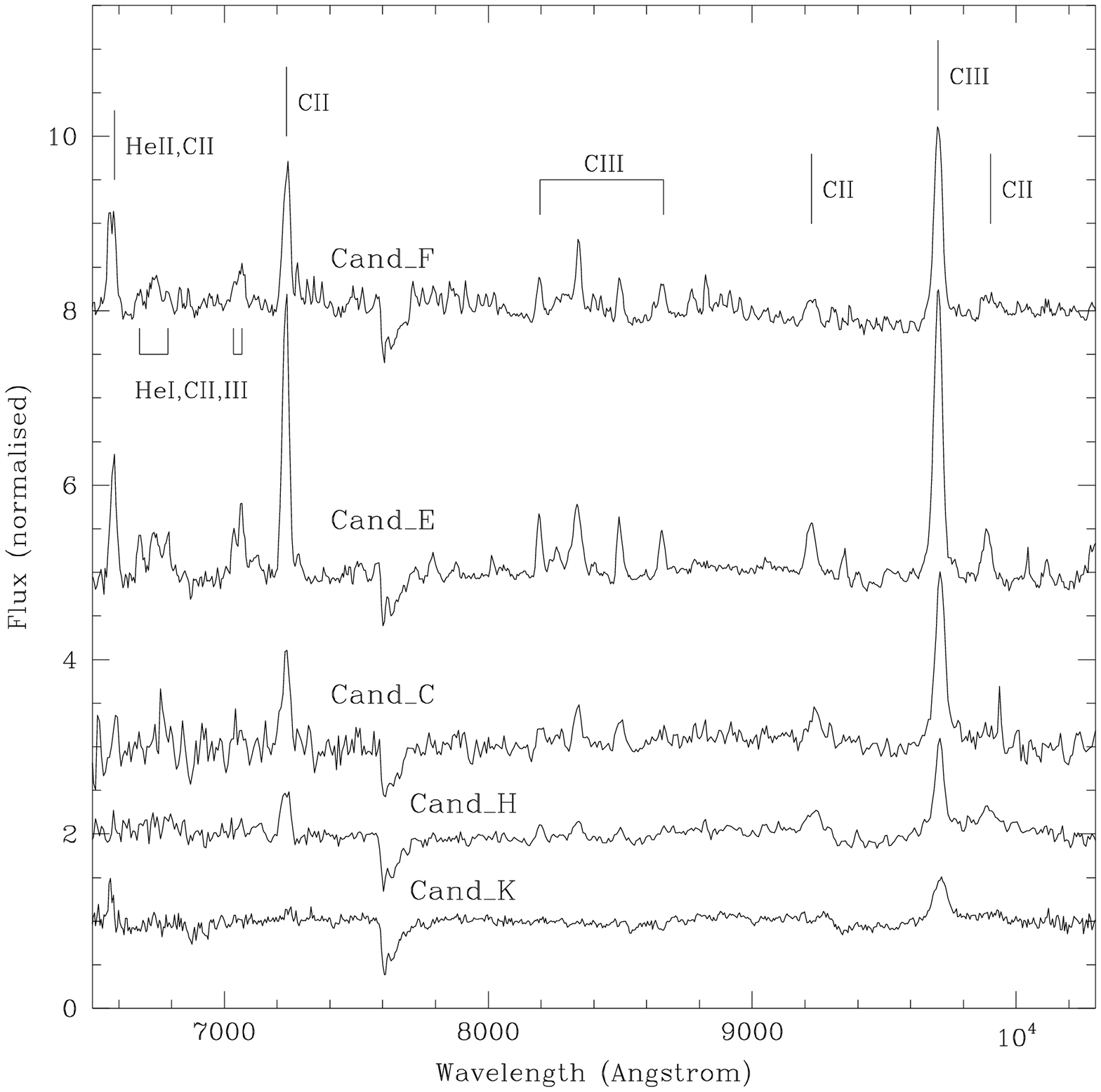}{11cm}{0}{60}{60}{-200}{-100}
\caption{Spectra of  the newly discovered WC candidates in Wd\,1,
with prominent transitions identified.}
\end{figure}

Spectral classification of WR stars can only be approximate, since the main
spectral indicators lie on the blue region.  For the WC candidates,
we use the ratio of the
C\,{\sc iii}\,8\,500\,{\AA}/C\,{\sc iv}\,8\,856\,{\AA} and C\,{\sc
ii}\,9\,900\,{\AA}/C\,{\sc iii}\,9\,710\,{\AA} lines. All candidates
appear to be late, with candidates E, F and H likely to be WC9 stars, while
candidate C is likely to be WC8.
Among the WN stars, the broad-lined candidate A is likely to be WN4-5,
with candidates B and J looking WNE and candidates D, G, and I looking
WNL (WN6-8).

\vspace*{-2mm}
\section{The supergiant population}
\vspace*{-2mm}

A significant number of the spectra obtained seem to correspond to
high-lumino\-si\-ty stars, judging from the intensity of traditional
luminosity indicators, such as the O\,{\sc i}\,7\,774\,\AA\ line or the
Ca\,{\sc ii} triplet in the $I$-band.

In particular, four stars with A or G spectral types have equivalent widths
for O\,{\sc i}\,7\,774\AA\ rather larger than Ia supergiants of similar
spectral types, confirming the result of Westerlund (1987).  These are
stars Wd\,1-4, Wd\,1-12, Wd\,1-16 and Wd\,1-265 (following the notation of Westerlund
1987).  All four stars are detected as relatively bright radio sources
(Dougherty \ea in preparation).  At least two other yellow supergiants are
of rather high luminosity.  From the very approximate classifications that
we can derive from our low-resolution spectra, we find that the total
number of A-G supergiants approaches the dozen, with several other stars
likely to be luminous B-type supergiants.  Many of these objects display
H$\alpha$ in emission.

In addition, there are four objects whose spectra resemble high-luminosity
M-type stars.  Three of them (Wd\,1-20, Wd\,1-26 and Wd\,1-237) are clearly detected
as radio sources.  In particular, Wd\,1-26 seems to be associated with
extended two-lobed radio emission (Clark \ea 1998).  The optical spectrum
of this object displays several emission lines unexpected in an M-type
star.

A second object associated with extended radio emission is Wd\,1-9.  Its
optical spectrum is characterised by a blue continuum with many strong
emission lines corresponding to both permitted and forbidden transitions.
This object is likely a Luminous Blue Variable.

\begin{table}
\caption{Coordinates for the newly identified Wolf-Rayet stars
in Westerlund\,1 determined  from 3.6-cm radio images (Dougherty \ea in
preparation).
Errors in the coordinates are $\sigma_\alpha$\,=\,$\pm$\,0.$^{\rm s}003$
and $\sigma_\delta$\,=\,$\pm$\,$0\farcs04$. For three of the objects (A, E and F)
Westerlund (1987) provides photographic magnitudes.
}
\begin{center}
\small
\begin{tabular}{ccllccc}
\hline \hline
          &        &        &          &         &                  &                                                      \\[-2mm]
WR        & Wd\,1  & WR$^a$ & spectral &   $V$   & R.A. (J2000)     &  Dec. (J2000)                                        \\
candidate & no.    &        & type     &  (phot) &                  &                                                      \\
          &        &        &          &         &                  &                                                      \\[-2mm]
\hline
          &        &        &          &         &                  &                                                      \\[-2mm]
J         &        & 77a    & WNE      & $>$\,19 & 16$^h$\,47$^m$\,0.$^s$885 & --\,45$^{\circ}$\,51$^{\prime}$\,20\farcs85 \\
I         &        & 77b    & WNL-8    & $>$\,19 & 16$^h$\,47$^m$\,1.$^s$668 & --\,45$^{\circ}$\,51$^{\prime}$\,20\farcs40 \\
K         &        & 77c    & WN       & $>$\,19 & 16$^h$\,47$^m$\,2.$^s$697 & --\,45$^{\circ}$\,50$^{\prime}$\,57\farcs35 \\
H         &        & 77d    & WC9      & $>$\,19 & 16$^h$\,47$^m$\,3.$^s$905 & --\,45$^{\circ}$\,51$^{\prime}$\,19\farcs88 \\
G         &        & 77e    & WN6-8    & $>$\,19 & 16$^h$\,47$^m$\,4.$^s$015 & --\,45$^{\circ}$\,51$^{\prime}$\,25\farcs15 \\
C         &        & 77f    & WC8      & $>$\,19 & 16$^h$\,47$^m$\,4.$^s$395 & --\,45$^{\circ}$\,51$^{\prime}$\,03\farcs79 \\
F         & 239    & 77g    & WC9      &  17.05  & 16$^h$\,47$^m$\,5.$^s$213 & --\,45$^{\circ}$\,52$^{\prime}$\,24\farcs97 \\
B         &        & 77h    & WNE      & $>$\,19 & 16$^h$\,47$^m$\,5.$^s$354 & --\,45$^{\circ}$\,51$^{\prime}$\,05\farcs03 \\
E         & 241    & 77i    & WC9      &  17.17  & 16$^h$\,47$^m$\,6.$^s$056 & --\,45$^{\circ}$\,52$^{\prime}$\,08\farcs26 \\
D         &        & 77j    & WN6-8    & $>$\,19 & 16$^h$\,47$^m$\,6.$^s$243 & --\,45$^{\circ}$\,51$^{\prime}$\,26\farcs48 \\
A         & 72     & 77k    & WN4-5    &  17.84  & 16$^h$\,47$^m$\,8.$^s$324 & --\,45$^{\circ}$\,50$^{\prime}$\,45\farcs51 \\
          &        &        &          &         &                  &                                                      \\[-2mm]
\hline\hline
\end{tabular}
\end{center}

Note $a$: WR number in the catalogue system of van der Hucht (2001).
\end{table}
\normalsize

A second very luminous blue object (Wd\,1-243) presents moderately strong
emission lines and appears to be surrounded by nebulosity.  This object is
also a radio source.  Extended radio emission to the NW of the cluster
(Dougherty \ea in preparation) is associated with some very faint
emission-line objects.  Other blue stars display strong emission lines and
are likely to be different kinds of transitional evolved objects.

\vspace*{-2mm}
\section{Discussion}
\vspace*{-2mm}

We have detected eleven WR stars in Westerlund\,1, the largest number of WR
stars known in any Galactic cluster, with the possible exception of the
Arches cluster (Blum \ea 2001).  Moreover, our survey is very incomplete:
only about $25$\% of the stars at the apparent magnitude typical of the WR
stars detected have been observed.  There is an obvious lack of WR
detections in the central region of the cluster (where the most luminous
supergiants are located), which strongly suggests that our sample is also
affected by observational effects.  We can conservatively assume that the
actual WR population of Wd\,1 is easily twice as large.

Wd\,1 appears to be unique among Galactic clusters in both the large number
and variety of massive post-main sequence (PMS) objects.  Since published
determinations of distance and reddening to the cluster are inaccurate and
inconsistent (Westerlund 1987; Piatti, Bica \& Clari\'a 1998), and the
field is affected by strong and probably variable reddening, we are still
unable to provide accurate values for the intrinsic luminosity of members
from which good estimates of the cluster age can be derived.  Several lines
of argument, however, suggest that the cluster is potentially extremely
massive.

On one side, the lack of an identifiable MS turnoff in available
photometric data prevents us from determining cluster parameters, but at
the same time provides us with a clue to the cluster size, suggesting
that all observed members down to $V$\,$\simeq$\,18 (in excess of 90) are
evolved stars with intrinsic magnitudes in the M$_V$\,$\approx$\,--\,6 to
--\,10 range.

On the other hand, the large number of transitional objects observed in
short-lived phases is suggestive of a very large population from which they
are evolving.  In particular, the number of yellow hypergiants in Wd\,1
appears comparable to that in the rest of the Galaxy.  Given the short
duration of the hypergiant phase, a population of several hundred O-type
stars seems to be suggested.

The presence of a large number of yellow supergiants (and a few very
luminous red supergiants) indicates that Wd\,1 is slightly older than the
very massive Arches Cluster.  Assuming that the progenitors of the yellow
hypergiants had ZAMS masses of $\sim$\,40\,-\,50\,M$_\odot$ (which can be
considered an educated guess), the age of the cluster would be
$\sim$\,4\,-\,5\,Myr, compatible with the presence of a population of WC
stars descending from more massive progenitors.  The alternative of
considering that the progenitors of the yellow supergiants have lower
($\sim$\,30\,M$_\odot$) masses and taking an older age of 8\,-\,10\,Myr in
order to consider that the red stars are normal M-type supergiants seems to
conflict with the large population of WR stars if we consider that they are
mainly single stars.  A large population of WR stars at an age of $\sim 8$
Myr could, however, be possible if most of them are part of binary systems.

With the much extended data set available after our 2002 observing
campaign, we expect to be able to provide accurate determinations for the
distance and extinction to Wd\,1.  The lack of radio detections for most WR
stars (Dougherty \ea in preparation) suggests that the distance to Wd\,1
has to be $\ga$\,2\,kpc.  It cannot be, however, much larger than the
5\,kpc advanced by Westerlund (1987), specially since the results of Piatti
\ea (1998) suggest that the interstellar absorption (which has a very
important component local to the cluster region) may be higher than
estimated by Westerlund (1987).  The cluster is then likely to be located
in the Crux Arm, at a distance of 4\,-\,5\,kpc.

In any case, with a true distance modulus $DM$\,$<$\,14\,mag, Wd\,1 offers
the unrivalled opportunity of observing its whole stellar population using
existing instrumentation (at least, in the near-IR).  It therefore
represents an ideal laboratory for the study of the impact of the presence
of a large population of massive stars on its environment and,
specifically, on the formation of lower mass stars.

\vspace*{+2mm}
\acknowledgments
Based on observations collected at the ESO, La Silla, Chile (ESO
67.D-0211).  We thank Sean Dougherty and Rens Waters for access to their
data on Wd\,1 and many helpful discussions.  We would also like to thank
many of our colleagues for very informative discussions, in particular
Simon Goodwin, Richard Stothers, Paul Crowther, Kees de Jager and Hans
Nieuwenhuijzen.

\vspace{6mm}
\noindent
{\bf Discussion}
\vspace{3mm}

\small

\noindent
{\sc Crowther}: Your discovery that Wd1 contains (at least) 11 WR stars
plus RSGs and YHGs is truely remarkable.  Its low radio flux, together with
the fact that {\sl MSX} mid-infrared images reveal the local environment
evacuated around Wd1 surely point to an age loser to 10\,Myr than
4\,-\,5\,Myr.  \\

\noindent
{\sc Negueruela}: With available data, it is still very difficult to
provide much more than a guess for the cluster age.  While the presence of
several very luminous blue objects seems to favour the lower age, it is
true that several lines of argument indicate an age around 10\,Myr.  We
really need to proceed further with our analysis before we can give a more
consistant estimate of even decide on the overall co-evality of the
cluster.

\end{document}